\documentclass[draftclsnofoot,12pt, onecolumn]{IEEEtran}
\usepackage{amsmath}
\usepackage{epsfig}
\usepackage{amssymb}
\usepackage{latexsym}
\usepackage{graphicx}
\usepackage{cite}
\title{Novel Expressions for the Rice $Ie{-}$Function \\ and the Incomplete Lipschitz-Hankel Integrals}
\author{\IEEEauthorblockN{Paschalis C. Sofotasios\\}
\IEEEauthorblockA{School of Electronic and Electrical Engineering\\
University of Leeds, UK\\
e-mail: eenpso@leeds.ac.uk\\}
\and
\IEEEauthorblockN{Steven Freear\\}
\IEEEauthorblockA{School of Electronic and Electrical Engineering\\
University of Leeds, UK\\
e-mail: s.freear@leeds.ac.uk}}
\begin{document}
\maketitle
\begin{abstract} 
This paper presents novel analytic expressions for the Rice $Ie{-}$function, $Ie(k,x)$, and the incomplete Lipschitz-Hankel Integrals (ILHIs) of the modified Bessel function of the first kind, $Ie_{m,n}(a,z)$. Firstly, an exact infinite series and an accurate polynomial approximation are derived for the $Ie(k ,x)$ function which are valid for all values of $k$. Secondly, an exact closed-form expression is derived for the $Ie_{m,n}(a,z)$ integrals for the case that $n$ is an odd multiple of $1/2$ and subsequently an infinite series and a tight polynomial approximation which are valid for all values of $m$ and $n$. Analytic upper bounds are also derived for the corresponding  truncation errors of the derived series'. Importantly, these bounds are expressed in closed-form and are particularly tight while they straightforwardly indicate that a remarkable accuracy is obtained by truncating each series after a small number of terms. Furthermore, the offered expressions have a convenient algebraic representation which renders them easy to handle both analytically and numerically.  As a result, they can be considered as useful mathematical tools that can be efficiently utilized in applications related to the analytical performance evaluation of classical and modern digital communication systems over fading environments, among others.
\end{abstract}
$ $\\
%
%
\section{Introduction}
\indent
It is widely accepted that special functions constitute invaluable mathematical tools in almost all areas of natural sciences and engineering. In the wide field of digital communications they have been extensively used in numerous analytic studies where they typically constitute the derivation of explicit expressions for important performance measures possible. Such measures correspond to the output signal-to-noise-ratio (SNR), channel capacity, error probability and higher-order statistics, to name a few.  In addition, the usefulness of special functions has became ever more evident over the last decades due to the fact that the majority of them is included as built-in functions in popular scientific software packages such as  $Maple$, $Matlab$ and $Mathematica$. As a result, their traditionally laborious computational realisation as well as their algebraic complexity have been significantly simplified. \\
\indent
Among others, the Rice $Ie$-function and the incomplete Lipschitz-Hankel integrals (ILHIs) have been shown to be useful in telecommunications since they have appeared in various analytic solutions of problems related to wireless communication systems. They were both proposed a few decades ago and they are denoted as $Ie(k,x)$ and $ \mathcal{Z}e_{m,n}(a,z)$, respectively \cite{B:Abramowitz}. In more details, the $Ie(k,x)$ function is typically defined by a lower incomplete integral which involves an exponential function and a modified Bessel function of the first kind and zero order \cite{J:Rice}. Alternative algebraic representations include two series expressions which were reported in \cite{J:Tan}. These series are infinite and are represented in terms of the modified Struve function and the modified Bessel function of the first kind, \cite{B:Abramowitz, B:Tables}. Moreover, the author in \cite{J:Pawula} reported useful closed-form identities between the $Ie$-function and the Marcum Q-function \cite{J:Marcum, J:Swerling, B:Proakis, B:Alouini}. The usefulness of $I_{e}(k, x)$ function in digital communications is based on the fact that it has been employed in the study of zero crossing rates, in the analysis of angle modulation systems and in radar pulse detection and error rates in differentially encoded systems, \cite{J:Tan, J:Pawula, B:Roberts, J:Rice_1}. \\
\indent 
In the same context, the ILHIs belong to a class of incomplete cylindrical functions which are one of the $J_{\nu}(x)$, $I_{\nu}(x)$, $Y_{\nu}(x)$, $K_{\nu}(x)$ and $H_{\nu}(x)$ Bessel functions, namely: the Bessel function of the first kind, the modified Bessel function of the first kind, the Bessel function of the second kind, the modified Bessel function of the second kind and the Hankel functions, respectively \cite{B:Abramowitz, B:Tables}. In the wide field of electrical engineering, the ILHIs have been largely encountered in analytic solutions of numerous problems related  to the theory of electromagnetics, while in communication theory they have been used in recent investigations associated with the error rate analysis of Multiple-Input-Mutiple-Output (MIMO) systems under imperfect channel state information (CSI) employing adaptive modulation, transmit beamforming and maximal ratio combining (MRC), \cite{B:Maksimov, J:Dvorak, J:Paris} (and the references therein).  \\ 
\indent
Nevertheless, even thought the above functions have been proved particularly useful in digital communications, it is noticed that they are both neither available in tabulated form, nor are they included as built-in functions in the aforementioned popular scientific software packages. As a consequence, their corresponding analytical tractability as well as computational realisation appear to be rather laborious and cumbersome. Motivated by this, the aim of this work is the derivation of analytic expressions for the Rice $Ie(k, x)$ function and the $Ie_{m, n}(a, z)$ integrals. Specifically, a remarkably accurate closed-form approximation is firstly derived for the $Ie(k, x)$ function. Subsequently, an exact closed-form expression is derived for the $Ie_{m,n}(a,z)$ integrals for the special case that $n$ is an odd multiple of $1/2$, i.e. $n + 0.5 \in \mathbb{N}$ and a polynomial approximation which, as in the case of $Ie(k,x)$ function, is valid for all parametric values. Furthermore, upper bounds are derived for the corresponding truncation error of the proposed polynomial approximations. Importantly, these bounds are expressed in closed-form and are shown to be particularly tight. It is also noted that the derived expressions have a tractable algebraic representation which ultimately renders them easily computable and useful in various analytical studies in wireless communications. Indicatively, such studies include the derivation of explicit expressions for vital performance measures, such as channel capacity and probability of error, in the wide field of digital communications over fading channels and the information-theoretic analysis of MIMO systems. \\
\indent
The remainder of this paper is structured as follows: Section II revisits the definition and basic identities of the Rice $Ie$ function and the $Ie_{m, n}(a, z)$ integrals. Subsequently, Sections III and IV are devoted to the derivation of novel analytic expressions for each of these functions, respectively, Finally, discussion on the potential applicability of the offered relationships in wireless communications along with closing remarks, are provided in Section V. 
%
%
\section{Definitions and existing representations.}
\subsection{The Rice $Ie$-function} 

\textbf{Definition 1.} \textit{For $k, x \in \mathbb{R^{+}}$ and $0 \leq k \leq 1$, the Rice $Ie$-function is defined as,}

\begin{equation} \label{1}
Ie(k,x) = \int_{0}^{x} e^{-t} I_{0}(kt) dt, 
\end{equation}
\textit{where $I_{0}(.)$ is the modified Bessel function of the first kind and zero order, } \cite{J:Tan, J:Pawula, B:Abramowitz, B:Tables}. \\ 
$ $\\
An equivalent integral representation to \eqref{1} was given in \cite{J:Tan}, namely,

\begin{equation} \label{2}
Ie(k,x) = \frac{1}{\sqrt{1 - k^{2}}} - \frac{1}{\pi} \int_{0}^{\pi} \frac{e^{-x(1 - cos\theta)}}{1 - k cos\theta} d\theta ,
\end{equation}
along with the following alternative series expressions, 
 
\begin{equation} \label{3}
Ie(k,x) = \sqrt{\frac{x \pi}{2 \sqrt{1 - k^{2}}}} e^{-x} \sum_{n = 0}^{\infty} \frac{1}{n!} \frac{x^{n} k^{2n}}{2^{n} \sqrt{1 - k^{2}}}  \left[ \frac{1}{\sqrt{1 - k^{2}}} L_{n + \frac{1}{2}} \left(x\sqrt{1 - k^{2}} \right) + L_{n - \frac{1}{2}} \left(x\sqrt{1 - k^{2}} \right) \right] 
\end{equation}
and

$$
Ie(k,x) = x e^{-x} \frac{\sqrt{\pi}}{2} \sum_{n = 0}^{\infty} \left( \frac{x\left(1 - k^{2} \right)}{2k} \right)^{n+1} \frac{I_{n+1}(kx)}{\Gamma \left(n + \frac{5}{2} \right)} \, +
$$
\begin{equation} \label{4}
x e^{-x} \left[I_{0}(kx) + \frac{\sqrt{\pi}}{2k} \sum_{n = 0}^{\infty} \left( \frac{x\left(1 - k^{2} \right)}{2k} \right)^{n} \frac{I_{n+1}(kx)}{\Gamma \left(n + \frac{3}{2} \right)} \right].
\end{equation}
The notations $L_{n}(.)$ and $\Gamma(.)$ denote the modified Struve function and the gamma function, respectively \cite{B:Abramowitz, B:Tables}. It is recalled that \eqref{3} converges relatively quickly for the case that $x\sqrt{1 - k^{2}}$ is large and $kx$ is small, whereas \eqref{4} converges quickly when $x\sqrt{1 - k^{2}}$ is small and $kx$ is large. Nevertheless, computing the $Ie(k,x)$ function using these expressions is inefficient due to the following three reasons: i) two relationships are required; ii) the above series are relatively unstable due to their infinite form; iii) the $L_{n}(.)$ function is not built-in in widely used mathematical software packages. \\
\indent
An adequate way of resolving this issue was reported in \cite{J:Pawula} where the $Ie(k, x)$ function is related to the known Marcum Q-function of the first order, namely, 

\begin{equation} \label{5}
Ie(k,x) = \frac{1}{\sqrt{1 - k^{2}}} \left[2Q(a , b) - e^{-x}I_{0}(kx) - 1 \right] 
\end{equation} 
and

\begin{equation} \label{6}
Ie(k,x) = \frac{1}{\sqrt{1 - k^{2}}} \left[ Q(a, b) - Q(b, a) \right]
\end{equation}
where,

\begin{equation}
a=\sqrt{x}\sqrt{1 + \sqrt{1 - k^{2}}},
\end{equation}

\begin{equation}
b=\sqrt{x}\sqrt{1 - \sqrt{1 - k^{2}}}
\end{equation}

and

\begin{equation} \label{7}
Q_{1}(a,b) = Q(a,b) \triangleq \int_{b}^{\infty}x e^{-\frac{x^{2} + a^{2}}{2}}I_{0}(ax)dx
\end{equation}
which was firstly proposed in \cite{J:Marcum}.
\subsection{The Incomplete Lipschitz-Hankel Integrals.}

\textbf{Definition 2.} \textit{For $z \in \mathbb{R^{+}}$ and $\forall \, a, m, n \in \mathbb{R} $, the generalized Incomplete Lipschitz-Hankel Integral (ILHI) is defined as,}

\begin{equation}  \label{8}
\mathcal{Z}e_{m,n}(a, z) = \int_{0}^{z}x^{m}e^{-ax} \mathcal{Z}_{n}(x)dx
\end{equation}
\textit{where $ \mathcal{Z}_{n}(x)$ denotes one of the cylindrical functions\footnote{The present analysis is limited to the consideration of the $I_{n}(x)$ function.} $J_{n}(x)$, $I_{n}(x)$, $Y_{n}(x)$, $K_{n}(x)$, $H_{n}^{1}(x)$ or $H_{n}^{2}(x)$,} \cite{B:Abramowitz, B:Tables}.\\
$ $\\
According to \cite{B:Maksimov, J:Dvorak} the parameters $m$, $n$, $a$, $z$ can be also complex.  An alternative representation for the ILHIs of the first-kind modified Bessel functions in terms of the Marcum Q-function was recently reported in \cite{J:Paris}, namely,

$$
Ie_{m,n}(a, x) = A_{m,n}^{0}(a) + e^{-ax} \sum_{i=0}^{m}\sum_{j=0}^{n+1}B_{m,n}^{i,j}(a)x^{i}I_{j}(x) 
$$
\begin{equation} \label{9}
+ A_{m,n}^{1}(a)Q_{1}\left(\sqrt{\frac{x}{a+\sqrt{a^{2}-1}}}, \sqrt{x}\sqrt{a+\sqrt{a^{2}-1}} \right)
\end{equation}
where the coefficients $A_{m,n}^{l}(a)$ and $B_{m,n}^{i,j}(a)$ can be obtained recursively. As already mentioned, the above expression was derived in the context of the error rate analysis of MIMO systems with imperfect channel state information (CSI). 
%
%
\section{Novel Expressions For the Rice $Ie$-Function.}
\indent 
This Section is devoted to the derivation of a tight polynomial approximation for the Rice $Ie$-function.\\
$ $ \\
\noindent
\textbf{Proposition 1.}  \textit{For $x, k \in \mathbb{R^{+}}$ and $0 \leq k \leq 1$, the following expression holds,}

\begin{equation} \label{10}
Ie(k, x) \simeq \sum_{l = 0}^{L} \frac{\Gamma(L+l) L^{1 - 2l} k^{2l}\gamma( 1 + 2l,x)}{l!\Gamma(L-l+1)\Gamma(l+1)2^{2l}} 
\end{equation}
\textit{where $\Gamma(x)$ and $\gamma(a,x)$ denote the complete and incomplete gamma functions, respectively. }

\textbf{\textit{Proof.}} It is firstly recalled that a simple and remarkably tight approximation for the $I_{n}(x)$ function was reported in \cite[eq. (20)]{J:Gross}, namely,

\begin{equation} \label{11}
I_{n}(x) \simeq \sum_{l=0}^{L} \frac{\Gamma(L+l)L^{1-2l}}{l!\Gamma(L - l +1)\Gamma(n+l+1)} \left(\frac{x}{2} \right) ^{n +2l}.
\end{equation}
This expression differs from the corresponding Taylor series representation by the following three terms: $L^{1 - 2l} \Gamma(L + l) / \Gamma(L - l +1)$ and has been shown to be more stable and better convergent. Therefore, by setting $n = 0$ and $x = kt$ in (11) and substituting in \eqref{1}, it follows that:

\begin{equation} \label{12}
Ie(k, x) \simeq \sum_{l = 0}^{L} \frac{\Gamma(L+l) L^{1 - 2l} k^{2l}}{l!\Gamma(L-l+1)\Gamma(l+1)2^{2l}}\int_{0}^{x} t^{2l}e^{-t}dt .
\end{equation}
Notably, the above integral can be expressed in closed-form in terms of the lower incomplete gamma function according to \cite[eq. (3.381.3)]{B:Tables}. As a result, equation (10) is deduced and hence, the proof is completed. $\blacksquare$ \\
$ $\\
\textbf{\textit{Remark.}} As $L \rightarrow \infty$, equation (10) degenerates into the following exact infinite series representation,

\begin{equation} \label{13}
Ie(k, x) = \sum_{l= 0}^{\infty} \frac{k^{2l}\gamma( 1 + 2l,x)}{l!  \Gamma(l+1)2^{2l}} 
\end{equation}
which to the best of the authors' knowledge has not been previously reported in the open technical literature. 
\begin{figure}[h]
\centerline{\psfig{figure=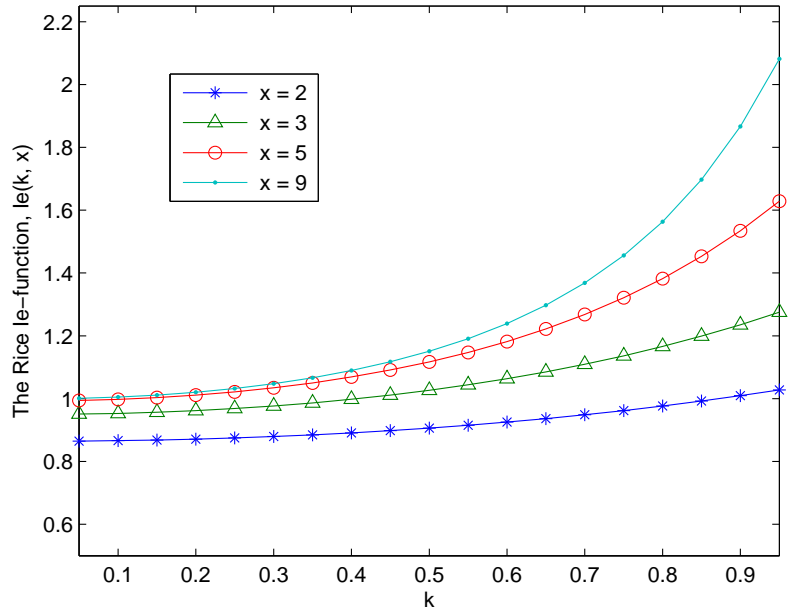, width=12cm, height=9cm}}
\caption{Behaviour of the polynomial approximation in (10) w.r.t $k$ for different values of $x$ and truncation after $20$ terms.}
\end{figure}
\subsection{A Closed-form Upper Bound for the Truncation Error.}
\indent
As already mentioned, equation (10) is an accurate polynomial approximation to the $Ie(k, x)$ function. Evidently, the corresponding accuracy increases with $L$ and as with all cases involving non-finite series expansions, a closed-form bound for the truncation error is undoubtedly necessary. 
\begin{figure}[h]
\centerline{\psfig{figure=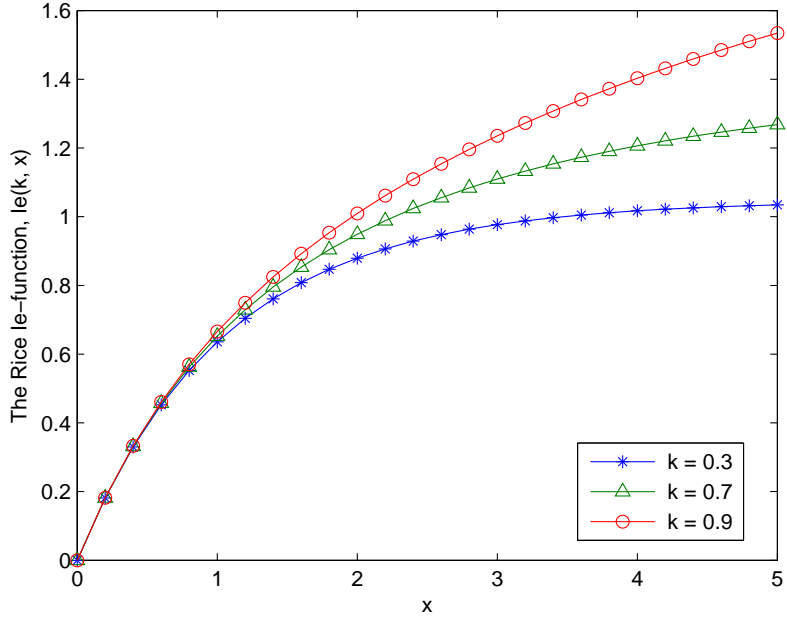, width=12cm, height=9cm}}
\caption{Behaviour of the polynomial approximation in (10) w.r.t $x$ for different $k$ values and truncation after $20$ terms.}
\end{figure}
$ $ \\
\\
\noindent
\textbf{Lemma 1.}  \textit{For $k, x \in \mathbb{R^{+}}$ and $0 \leq k \leq 1$,  the following inequality holds,}

\begin{equation} \label{14}
\epsilon_{t} <  1  + \sqrt{\frac{k}{2}} \left[\frac{erf(\sqrt{x}\sqrt{1-k})}{\sqrt{1 - k}} - \frac{erf(\sqrt{x}\sqrt{1+k})}{\sqrt{1+k}} \right] - I_{0}(kx)e^{-x} - \sum_{l = 0}^{L} \frac{\Gamma(L+l) L^{1 - 2l} k^{2l}\gamma( 1 + 2l,x)}{l!\Gamma(L-l+1)\Gamma(l+1)2^{2l}} 
\end{equation}
\textit{where erf(x) denotes the error function. }
\\
$ $ \\
\noindent
\textbf{\textit{Proof.}} For the case that (10) is truncated at the $L^{th}$ term, the corresponding truncation error is expressed as

\begin{equation} \label{15}
\epsilon_{t} = \sum_{l = L+1}^{\infty} \frac{\Gamma(L+l) L^{1 - 2l} k^{2l}\gamma( 1 + 2l,x)}{l!\Gamma(L-l+1)\Gamma(l+1)2^{2l}}. 
\end{equation} 
Since as already shown for $L \rightarrow \infty$, equation \eqref{10} yields an exact infinite series for the $Ie$-function, it follows straightforwardly that

\begin{equation} \label{16}
\epsilon_{t} = \sum_{l= 0}^{\infty} \frac{k^{2l}\gamma( 1 + 2l,x)}{l!  \Gamma(l+1)2^{2l}} - \sum_{l = 0}^{L} \frac{\Gamma(L+l) L^{1 - 2l} k^{2l}\gamma( 1 + 2l,x)}{l!\Gamma(L-l+1)\Gamma(l+1)2^{2l}} 
\end{equation}
which can be equivalently written as

\begin{equation} \label{17}
\epsilon_{t} = Ie(k, x)-  \sum_{l = 0}^{L} \frac{\Gamma(L+l) L^{1 - 2l} k^{2l}\gamma( 1 + 2l,x)}{l!\Gamma(L-l+1)\Gamma(l+1)2^{2l}}. 
\end{equation}
Importantly, the above expression can be upper bounded by using the tight closed-form upper bound for the $Ie(k, x)$ function reported in \cite[eq. (13)]{C:Sofotasios}, namely,

\begin{equation} \label{18} 
Ie(k, x) <  1 - e^{-x}I_{0}(kx) + \sqrt{\frac{k}{2}} \left[\frac{erf(c\sqrt{x})}{c} - \frac{erf(d\sqrt{x})}{d} \right]
\end{equation}
where $c = \sqrt{1 - k}$ and $d = \sqrt{1 + k}$. Hence, by substituting (18) into (17) yields (14), which completes the proof. $\blacksquare$\\
$ $\\
\textbf{\textit{Remark.}} By setting $L \rightarrow \infty$ in the the finite series in (14) yields a similar closed-form upper bound can be also obtained for the exact infinite series in (13), namely,

\begin{equation} \label{19}
\epsilon_{t} <  1+ \sqrt{\frac{k}{2}} \left[\frac{erf(\sqrt{x}\sqrt{1-k})}{\sqrt{1 - k}} - \frac{erf(\sqrt{x}\sqrt{1+k})}{\sqrt{1+k}} \right]  - I_{0}(kx)e^{-x}  - \sum_{l = 0}^{L} \frac{ k^{2l}\gamma( 1 + 2l,x)}{l! \Gamma(l+1)2^{2l}}. 
\end{equation}
\subsection{Numerical Results}
This Section is devoted to the validation and analysis of the performance of the offered results. To this end, Figure $1$ illustrates the behaviour of $Ie(k,x)$ function as function of $k$ for different values of $x$ and with the series truncated after $20$ terms. One can see the monotonically increasing behaviour of the function with respect to $k$. Notably, the series convergences particularly quickly since for all scenarios with the corresponding truncation after $20$ terms, the both the involved absolute error $\epsilon_{a} \triangleq \mid Ie(k, x) - \hat{I}e(k, x) \mid$ and the absolute relative error $\epsilon_{ar} \triangleq \mid Ie(k, x) - \hat{I}e(k, x) \mid / Ie(k, x) $ is significantly less than $10^{-10}$. \\
\indent
Figure $2$ depicts the behaviour of (10) as a function of $x$ for different values of $k$ and for truncation after $20$ terms. In the small value regime, the difference between the three scenarios is rather small; however, as $x$ increases the monotonically increasing behaviour of the function becomes more evident. Furthermore, the involved absolute and absolute relative errors are also very low, $\epsilon_{a}, \epsilon_{ar}<10^{-9}$, in every possible scenario.  
\begin{figure}[h]
\centerline{\psfig{figure=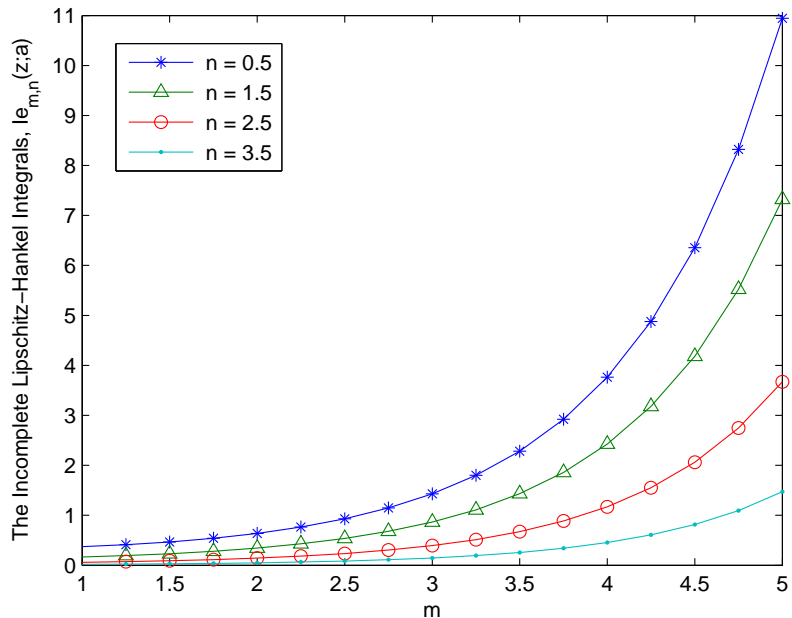, width=12cm, height=9.0cm}}
\caption{Behaviour of the closed-form expression in (20) w.r.t $m$ for $z = 4$, $a = 1.8$ and different values of $n$.}
\end{figure}
%
%
\section{Novel Expressions For the ILHIs.}
\indent 
This Section is devoted to the derivation of a closed-form expression for the case that $n$ is a half integer as well as a tight polynomial approximation which is valid for all values of the involved parameters.  
\subsection{A Closed-Form Expression for the case that $n + \frac{1}{2} \in \mathbb{N}$.}
$ $ \\
\noindent
\textbf{Theorem 1.}  \textit{For $a, m, z \in \mathbb{R}$, $a>1$ and $n + \frac{1}{2} \in \mathbb{N}$, the following expression holds,}

\begin{equation} \label{20}
Ie_{m,n}(a, z) = \sum_{k=0}^{n - \frac{1}{2}} \frac{\left(n+k-\frac{1}{2} \right)!}{\sqrt{\pi}k!\left(n-k-\frac{1}{2} \right)!2^{k+\frac{1}{2}}}  \left\lbrace (-1)^{k} \frac{\gamma \left(\mathcal{A}, (a-1)z \right) }{(a-1)^{\mathcal{A}}} + (-1)^{n + \frac{1}{2}} \frac{\gamma \left(\mathcal{A}, (a+1)z \right) }{(a+1)^{\mathcal{A}}} \right\rbrace
\end{equation}
\textit{where $\mathcal{A}=m-k+\frac{1}{2}$. }
\\
$ $ \\
\noindent
\textbf{\textit{Proof.}} It is recalled that for the case that $n + 0.5 \in \mathbb{N}$, the modified Bessel function of the first kind is expressed in closed-form by \cite[eq. (8.467)]{B:Tables}, namely, 

\begin{equation} \label{21}
I_{n + \frac{1}{2}}(x)\triangleq \sum_{k=0}^{n}\frac{(n+k)!\,\left[(-1)^{k}e^{x} + (-1)^{n+1}e^{-x}\right]}{\sqrt{\pi}k!(n-k)!(2x)^{k+\frac{1}{2}}}
\end{equation}
To this effect, by making the necessary change of variables and substituting in  (8), one obtains,

\begin{equation} \label{22}
Ie_{m,n}(a, z) = \sum_{k=0}^{n-\frac{1}{2}} \frac{\left(n+k- \frac{1}{2}\right)! 2^{-k - \frac{1}{2}}}{\sqrt{\pi} k!\left(n-k-\frac{1}{2} \right)!}  \left\lbrace (-1)^{k}\int_{0}^{z}x^{\mathcal{A}}e^{-ax}e^{x}dx + (-1)^{n + \frac{1}{2}} \int_{0}^{z} x^{\mathcal{A}}e^{-ax}e^{-x}dx \right\rbrace
\end{equation}
Notably, the integrals in (22) belong to the family of gamma special functions. Therefore, after some basic algebraic manipulation and with the aid of \cite[eq. (3.381.3)]{B:Tables}, equation (20) is deduced and therefore, the proof is completed. $\blacksquare$ 
\subsection{A Novel Series Representation}
$ $\\
\textbf{Proposition 2.}  \textit{For $m, a, n, z \in \mathbb{R}$, the following expression holds,}
$ $\\
\begin{equation} \label{23}
Ie_{m,n}(a, z)  \simeq \sum_{l=0}^{L} \frac{\Gamma(L+l)L^{1 - 2l}\gamma(m + n + 2l +1, az)}{l!(L-l)!\Gamma(n + l + 1)2^{n + 2l} a^{m+n+2l+1}}
\end{equation}
\textit{where $\Gamma(x)$ and $\gamma(a,x)$ denote the complete and incomplete gamma functions, respectively. }
\\
$ $ \\
\noindent
\textbf{\textit{Proof.}} Recalling the Gross' polynomial approximation for the $I_{n}(x)$ function, substituting (11) in (8), recalling that $a! \triangleq \Gamma(a+1)$ and performing some basic algebraic manipulations, one obtains 
\begin{figure}[h]
\centerline{\psfig{figure=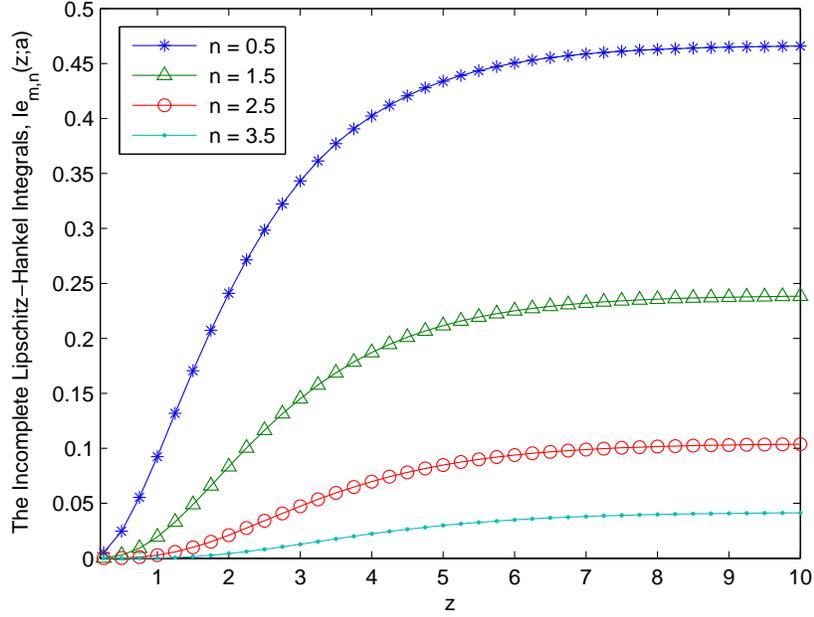, width=12cm, height=9cm}}
\caption{Behaviour of the closed-form expression in (21) w.r.t $z$ for $a = 1.8$, $m = 1.2$ and different values of $n$.}
\end{figure}

\begin{equation} \label{24}
Ie_{m,n}(a, z)  \simeq  \sum_{l=0}^{L} \frac{\Gamma(L+l)L^{1 - 2l} 2^{-n - 2l} }{l!(L-l)!\Gamma(n + l + 1)a^{m+n+2l+1}} \int_{0}^{z} x^{m+n+2l} e^{-ax}dx
\end{equation}
By making the necessary change of variables and using \cite[eq. (3. 381.3)]{B:Tables} yields (23) and thus, the proof is completed. $\blacksquare$ 
$ $\\
\\
\textbf{\textit{Remark.}} By recalling that (11) becomes exact as $L \rightarrow \infty$, it follows straightforwardly that $Ie_{m,n}(a, z) $ can be expressed by the following explicit expression,

\begin{equation} \label{25}
Ie_{m,n}(a, z)  = \sum_{l=0}^{\infty} \frac{ \gamma(m + n + 2l +1, az)}{l! \Gamma(n + l + 1)2^{n + 2l} a^{m+n+2l+1}}.
\end{equation}
To the best of the authors' knowledge the above expressions are novel.

\subsection{A Closed-Form Bound for the Truncation Error}

A closed-form bound for the corresponding truncation error can be derived by following the same methodology as in \textit{Lemma }$1$. \\
$ $ \\
\noindent
\textbf{Lemma 2.}  \textit{For $m, a, n, z \in \mathbb{R}$, the following inequality holds,}

$$
\epsilon_{t} < \sum_{k=0}^{\lfloor n - \frac{1}{2}\rfloor +\frac{1}{2}} \frac{\left(k + \lfloor n-\frac{1}{2}\rfloor + \frac{1}{2}\right)!}{\sqrt{\pi}k!\left(\lfloor n-\frac{1}{2}\rfloor +\frac{1}{2} - k\right)!2^{k+\frac{1}{2}}} \left\lbrace (-1)^{k} \frac{\gamma \left(\mathcal{A}, (a-1)z \right) }{(a-1)^{\mathcal{A}}} + (-1)^{\lfloor n + \frac{1}{2}\rfloor} \frac{\gamma \left(\mathcal{A}, (a+1)z \right) }{(a+1)^{\mathcal{A}}} \right\rbrace
$$
\begin{equation} \label{26}
 - \sum_{l=0}^{L} \frac{\Gamma(L+l)L^{1 - 2l}\gamma(m + n + 2l +1, az)}{l!(L-l)!\Gamma(n + l + 1)2^{n + 2l} a^{m+n+2l+1}}
 \end{equation}
\textit{where $\mathcal{A}=m-k+\frac{1}{2}$. }
\begin{figure}[h]
\centerline{\psfig{figure=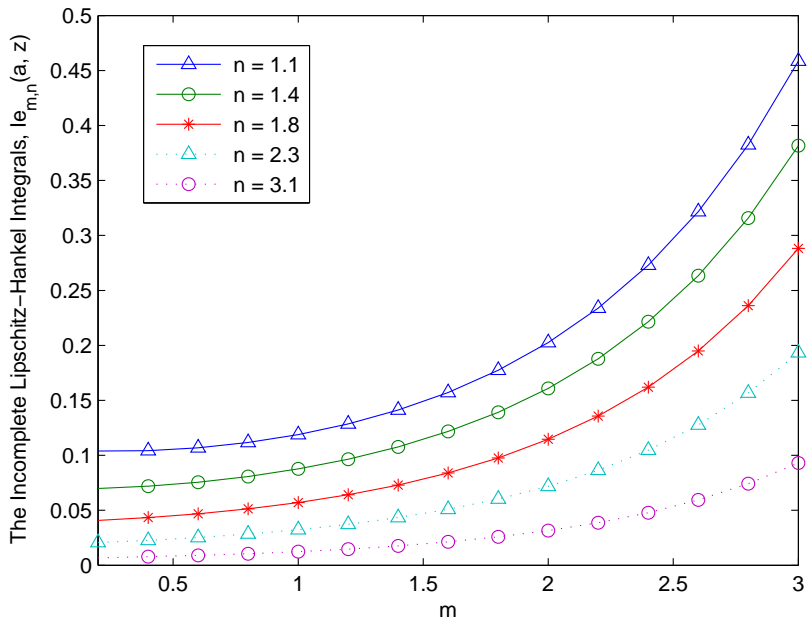, width=12cm, height=9cm}}
\caption{Behaviour of the closed-form expression in (23) w.r.t $m$ for $a = 2.2$, $z = 5$ and different values of $n$ for truncation after $700$ terms.}
\end{figure}
$ $ \\
\\
\noindent
\textbf{\textit{Proof.}} By truncating the series in (23) after $L$ terms, the corresponding truncation error is given by

\begin{equation} \label{27}
\epsilon_{t} = \sum_{l=L+1}^{\infty} \frac{\Gamma(L+l)L^{1 - 2l}\gamma(m + n + 2l +1, az)}{l!(L-l)!\Gamma(n + l + 1)2^{n + 2l} a^{m+n+2l+1}}
\end{equation}
Importantly, given that

\begin{equation}
\sum_{l=L+1}^{\infty} f(x) = \sum_{l = 0}^{\infty}f(x) - \sum_{l}^{L}f(x)
\end{equation}

and based on the derived series in (25), equation (27) is equivalently re-written as

\begin{equation} \label{28}
\epsilon_{t} = Ie_{m,n}(a, z)  - \sum_{l=0}^{L} \frac{\Gamma(L+l)L^{1 - 2l}\gamma(m + n + 2l +1, az)}{l!(L-l)!\Gamma(n + l + 1)2^{n + 2l} a^{m+n+2l+1}}
\end{equation}
It is recalled here that the $Ie_{m,n}(a, z)$ integral is monotonically decreasing with respect to $n$. Thus, by assuming an arbitrary positive real scalar $\mathfrak{b} \in \mathbb{R^{+}}$, the following inequality holds,

\begin{equation} \label{29}
Ie_{m,n+\mathfrak{b}}(a, z) < Ie_{m,n}(a, z)
\end{equation}
and thus $Ie_{m,n+\frac{1}{2}}(a, z) < Ie_{m,n}(a, z)  $. Therefore, by recalling the closed-form representation in \textit{Theorem $1$} and making use of the floor function principle in order to formulate the identity $n > \lfloor n - 0.5 \rfloor + 0.5$, the $Ie_{m,n}(a, z)$ function can be upper bounded $\forall n \in \mathbb{R^{+}} $ as follows,

$$
Ie_{m,n}(a, z) < \sum_{k=0}^{\lfloor n - \frac{1}{2}\rfloor +\frac{1}{2}} \frac{\left(k + \lfloor n-\frac{1}{2}\rfloor + \frac{1}{2}\right)!}{\sqrt{\pi}k!\left(\lfloor n-\frac{1}{2}\rfloor +\frac{1}{2} - k\right)!2^{k+\frac{1}{2}}} \times
$$
\begin{equation} \label{30}
 \left\lbrace (-1)^{k} \frac{\gamma \left(\mathcal{A}, (a-1)z \right) }{(a-1)^{\mathcal{A}}} + (-1)^{\lfloor n + \frac{1}{2}\rfloor} \frac{\gamma \left(\mathcal{A}, (a+1)z \right) }{(a+1)^{\mathcal{A}}} \right\rbrace
\end{equation}
where $\mathcal{A}=m-k+\frac{1}{2}$.
\indent
Therefore, by inserting (30) into (28) yields (26) and therefore completes the proof. $\blacksquare$
$ $ \\
\\
\textbf{\textit{Remark.}} By recalling that (23) becomes exact as $L \rightarrow \infty$, a corresponding closed-form bound for the truncation error of (25) is straightforwardly deduced, namely, 
\begin{figure}[h]
\centerline{\psfig{figure=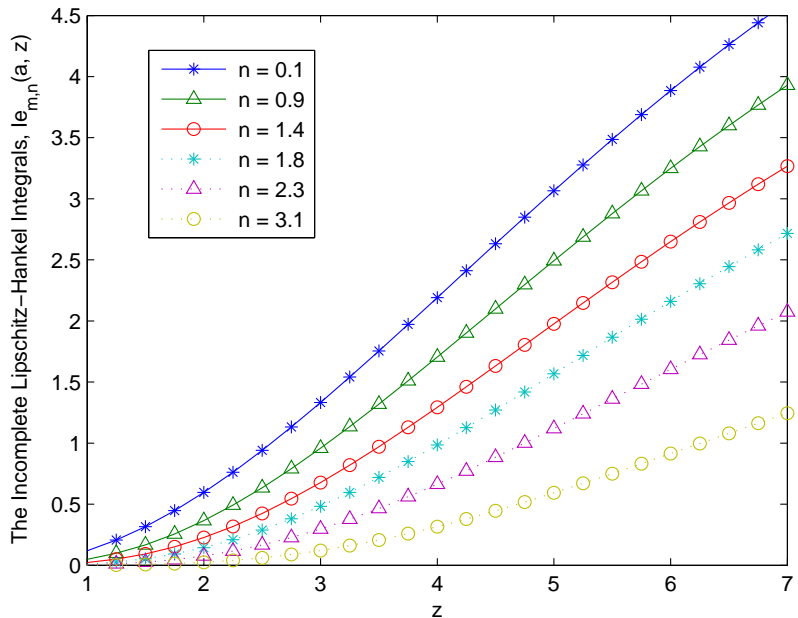, width=12cm, height=9cm}}
\caption{Behaviour of the polynomial approximation in (23) w.r.t $z$ for $a = 1.4$, $m = 2.2$ and different values of $n$ for truncation after $700$ terms.}
\end{figure}
 
$$
\epsilon_{t} < \sum_{k=0}^{\lfloor n - \frac{1}{2}\rfloor} \frac{\left(k + \lfloor n-\frac{1}{2}\rfloor \right)!}{\sqrt{\pi}k!\left(\lfloor n-\frac{1}{2}\rfloor - k\right)!2^{k+\frac{1}{2}}}\left\lbrace (-1)^{k} \frac{\gamma \left(\mathcal{A}, (a-1)z \right) }{(a-1)^{\mathcal{A}}} + (-1)^{\lfloor n + \frac{1}{2}\rfloor} \frac{\gamma \left(\mathcal{A}, (a+1)z \right) }{(a+1)^{\mathcal{A}}} \right\rbrace
$$
 
\begin{equation} \label{31}
 - \sum_{l=0}^{L} \frac{           \gamma(m + n + 2l +1, az)}{l!      \Gamma(n + l + 1)2^{n + 2l} a^{m+n+2l+1}}
\end{equation}
where again $\mathcal{A}=m-k+\frac{1}{2}$.

It is recalled that the offered results can be used in numerous areas in natural sciences and engineering including wireless communications with emphasis in digital communications over fading channels \cite{Paschalis_1, Paschalis_2, Paschalis_3, Paschalis_4, Paschalis_5, Paschalis_6} and the references therein. 
\subsection{Numerical Results} 
\indent
The validity of the derived closed-form expression and the general behaviour of the offered bounds are illustrated in Figures $3-6$. Figure $3$ depicts the behaviour of the closed-form expression in (20) with respect to $m$ for different half integer values of $n$ and assuming that $z = 4$ and $a = 1.8$. One can observe and verify the monotonically decreasing behaviour of the function with respect to $n$. This is also observed in Figure $4$ where the behaviour of (20) is depicted with respect to $z$ for $a=1.8$ and $m=1.2$. In the same context, Figures $5$ and $6$ illustrate the behaviour of the $Ie_{m,n}(a, z)$ in (23) for the case of arbitrary values of $n$ and for truncation after $700$ terms. The former is represented w.r.t $m$ for $a=2.2$ and $z = 5$ while the latter is plotted w.r.t $z$ for $a=1.2$ and $m=2.2$. The usefulness of the offered expressions is evident since (20) is an exact closed-form expression while (23) is simple to use and very accurate. It is additionally useful for its capability to account for the sensitivity of $n$.

\section{Conclusion}
\indent 
In this work, novel analytic results were presented for the Rice $Ie$-function and the Incomplete Lipschitz Hankel Integrals of the modified Bessel function of the first kind. The offered results have a convenient algebraic representation and therefore they are easy to handle both analytically and numerically. This is sufficiently advantageous since it constitutes them suitable for utilization in various studies related to the analytical performance evaluation of of digital communications over fading channels, among others. \\
$ $\\
\bibliographystyle{IEEEtran}

\end{document}